\documentclass[useAMS,usenatbib,fleqn]{mn2e}

\usepackage{amsmath}
\usepackage{amssymb}
\usepackage[utf8]{inputenc}
\usepackage{xspace}
\usepackage[american]{babel}
\usepackage{times}
\usepackage{graphicx}
\usepackage{color}
\usepackage{multirow}

\newcommand{\ie}{i.\,e.\xspace}

\newcommand{\eg}{e.\,g.\xspace}
 
\newcommand{\GRB}{Gamma-Ray Burst\xspace}
\newcommand{\GRBs}{Gamma-Ray Bursts\xspace}
\newcommand{\tv}{\theta_\mathrm{v}}
\newcommand{\tc}{\theta_\mathrm{c}}
\newcommand{\tj}{\theta_\mathrm{jet}}
\newcommand{\Eiso}{E_\mathrm{iso}}
\newcommand{\Liso}{L_\mathrm{iso}}
\newcommand{\Ep}{E_\mathrm{peak}}
\newcommand{\dex}{\thinspace\mathrm{dex}}
\newcommand{\erg}{\thinspace\mathrm{erg}}
\newcommand{\keV}{\thinspace\mathrm{keV}}
\newcommand{\tgtc}{\theta_\Gamma/\tc}

\newcommand{\simpropto}{\mathrel{\vcenter{
  \offinterlineskip\halign{\hfil$##$\cr
    \propto\cr\noalign{\kern2pt}\sim\cr\noalign{\kern-2pt}}}}}

\title[Structure of GRB jets: intrinsic \textit{versus} apparent]{Structure of \GRB jets: intrinsic \textit{versus} apparent properties}
\author[O. S. Salafia, G. Ghisellini, A. Pescalli, G. Ghirlanda, F. Nappo]{O. S. Salafia$^{1,3}$\thanks{E-mail:
omsharan.salafia@brera.inaf.it (OA Brera Merate), o.salafia@campus.unimib.it (Univ. Milano-Bicocca)}, G. Ghisellini$^{3}$, A. Pescalli$^{2,3}$, G. Ghirlanda$^{3}$, F. Nappo$^{2,3}$\\
$^{1}$Universit\`a degli Studi di Milano-Bicocca, Piazza della Scienza 3, I-20126 Milano, Italy\\
$^{2}$Universit\`a degli Studi dell'Insubria, Via Valleggio, 11, I-22100 Como, Italy \\
$^{3}$INAF - Osservatorio Astronomico di Brera Merate, via E. Bianchi 46, I–23807 Merate, Italy\\
}

\begin{document}
 
 \date{Draft version, \today}
 
 \pagerange{\pageref{firstpage}--\pageref{lastpage}} \pubyear{2015}
 
 \maketitle
 
 \label{firstpage}
 
\begin{abstract}
With this paper we introduce the concept of \textit{apparent structure} of a GRB jet, as opposed to its \textit{intrinsic} structure. The latter is customarily defined specifying the functions $\epsilon(\theta)$ (the energy emitted per jet unit solid angle) and $\Gamma(\theta)$ (the Lorentz factor of the emitting material); the apparent structure is instead defined by us as the isotropic equivalent energy $\Eiso(\tv)$ as a function of the viewing angle $\tv$. We show how to predict the apparent structure of a jet given its intrinsic structure. We find that a Gaussian intrinsic structure yields a power law apparent structure: this opens a new viewpoint on the Gaussian (which can be understood as a proxy for a realistic narrow, well collimated jet structure) as a possible candidate for a quasi-universal GRB jet structure. We show that such a model (a) is consistent with recent constraints on the observed luminosity function of GRBs; (b) implies fewer orphan afterglows with respect to the standard uniform model; (c) can break out the progenitor star (in the collapsar scenario) without wasting an unreasonable amount of energy; (d) is compatible with the explanation of the Amati correlation as a viewing angle effect; (e) can be very standard in energy content, and still yield a very wide range of observed isotropic equivalent energies.
\end{abstract}


\section{Introduction}

\begin{figure}
 \begin{center}
 \includegraphics{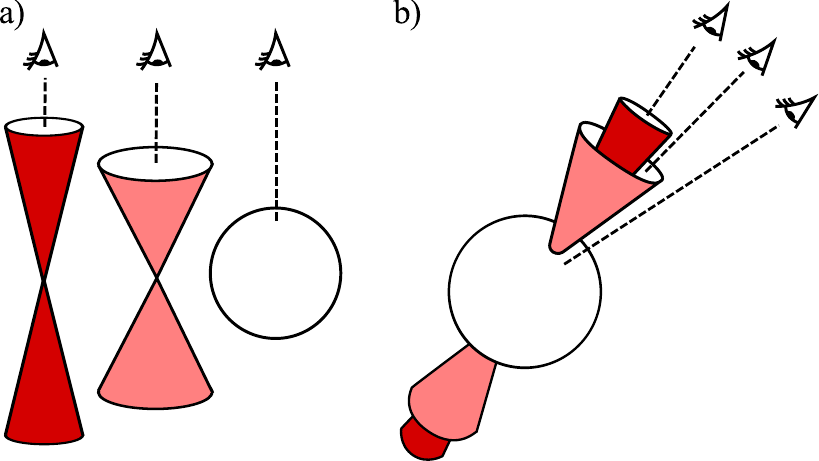}
 \caption{\label{fig:lipunov} Sketch of the two possible scenarios described in \citet{lipunov-grb_standard_energy2001}. In the first one (a) GRB jets are seen on-axis, and they differ by their semiaperture and consequently their observed energy, with the brightest being the most collimated; in the second one (b) the jet configuration is such that the viewing angle determines which component contributes most significantly to the received energy.}
 \end{center}
\end{figure}

In a 1999 preprint Lipunov, Postnov and Prokhorov introduced, possibly for the first time, the idea that \GRBs (GRBs) could be ``standard energy explosions'', \ie events with a standard energy reservoir. In their final article, published two years later \citep{lipunov-grb_standard_energy2001}, the authors identified $E_{0} \sim 5\times 10^{51}$ erg as a plausible value for this universal energy, and they described two possible scenarios: in the first, the standard energy is emitted inside a conical jet whose semiaperture $\tj$ varies from one GRB to another (Fig.~\ref{fig:lipunov}.a); in the second, the beam pattern, made up of two coaxial conical components and an isotropic component, is the same for all GRBs (Fig.~\ref{fig:lipunov}.b). In their view, the wide range of observed luminosities of GRBs could be accommodated in either the first scenario, with the brightest events being the most collimated, or the second picture, with the viewing angle being crucial to determine which part of the beam mostly contributes to the observed fluence.

Soon later \citet{Frail01} analyzed a sample of 17 Long GRBs for which a \textit{jet break} in the afterglow light curve was identified, and thus a measure of $\tj$ was available, finding a surprising clustering of the collimation-corrected energy $E_{\gamma}\equiv \Eiso (1-\cos\tj)\approx \Eiso\tj^2/2$ around the universal value $E_{\gamma} \sim 5\times 10^{50}$~erg (which implies a correlation $\Eiso\propto \tj^{-2}$). The result was interpreted as evidence that the emission is in fact beamed inside a conical jet: this supported the first scenario proposed by \citet{lipunov-grb_standard_energy2001}, finally tracing the very wide range of observed isotropic equivalent energies of GRBs to a single ``real'' value.

Next year, \citet{Rossi-structured2002} and \citet{zhang-universalconfig2002} interpreted the same result in a different way, closer to the second scenario proposed by \citet{lipunov-grb_standard_energy2001}: their claim was that the correlation $\Eiso\propto \tj^{-2}$ was instead due to the existence of a universal jet structure, with the jet energy per unit solid angle being $\epsilon(\theta)\propto \theta^{-2}$. This particular energy configuration, along with the assumption of a strong relativistic beaming of the emitted radiation, implies that $\Eiso\propto \tv^{-2}$, where $\tv$ is the viewing angle. Based on a simulation of the afterglow light curves produced by such a structured jet (SJ) \citet{Rossi-structured2002} claimed\footnote{In \citet{Rossi-structured2002}, the afterglow light curve of their SJ seen under a viewing angle $\tv$ is found to show a feature similar to the \textit{jet break} predicted for the uniform jet, with the coincidence that the break time is approximately the same as that of a uniform jet, seen on-axis, with semiaperture $\tj=\tv$. The correspondence is not exact, as discussed two years later in \citet{rossi-polarization-2004}, but the difference should be small in most cases.} that $\tv\sim\tj$, and thus $\Eiso\propto\tj^{-2}$.

The idea of a \textit{quasi-universal} jet structure (that is, universal with some dispersion of the structure parameters) stimulated a number of papers exploring the idea and its consequences. Here are some examples: \citet{granot-sj-afterglow-constraining-2003} and \citet{kumar-sj-afterglow-2003} constrained the possible jet structures by a qualitative comparison of simulated afterglow light curves with the observed ones; \citet{rossi-polarization-2004} showed that polarization measures could be a crucial tool to discern between the SJ and the uniform jet; \cite{2004ApJ...601L.119Z} and \cite{2005ApJ...621..875D} proposed the interpretation of X-Ray Flashes (XRFs) and GRBs as a unique population of bursts with a Gaussian structured jet, and tested this hypothesis against many observational constraints, finding general consistency; on the contrary, \cite{0004-637X-620-1-355} found that the universal structured jet proposed by \citet{Rossi-structured2002} fails to predict the right observed number ratio of XRFs to GRBs; relativistic hydrodynamical simulations (\eg \citet{2004ApJ...601L.119Z,morsony-grb-simula-2007}) showed that the interaction of the jet with the stellar envelope prior to the break out (in the collapsar scenario of Long GRBs) leads inevitably to some structure in the jet properties, but it remains unclear if this structure is likely to have any degree of universality; \cite{2013MNRAS.428.2430L}, within a photospheric emission model, obtained a low energy photon index $\alpha$ consistent with the observations assuming a SJ where the Lorentz factor $\Gamma(\theta)$ varies as a power law with the angular distance from the jet axis; we \citep{2015MNRAS.447.1911P} recently showed that the observed luminosity function of Long GRBs is consistent with the SJ model, provided that the energy structure is much steeper than the original $\theta^{-2}$ (at least $\theta^{-4}$ seems to be necessary).

Despite all these efforts and some successes, no consensus about the viability of the quasi-universal SJ hypothesis has been achieved so far.

\subsection{Aim and structure of this paper}

The aim of this article is to introduce the concept of \textit{apparent structure} and to show that it is a necessary tool to correctly compare the predictions of the SJ model with the observations. In the following sections, we will try to make the distinction between intrinsic and apparent structure as clear and as rigorous as possible; for the moment, suffice it to say that the intrinsic structure here is understood as the energy \textit{emitted} by different portions of the jet at different \textit{angular distances} $\theta$ from the jet axis, while the apparent structure describes the energy \textit{received} by observers that see the same jet under different \textit{viewing angles} $\tv$.
While the intrinsic structure can be due to the jet formation process itself (\eg the \citet{1977MNRAS.179..433B} process) or to the subsequent interaction of the jet with the stellar envelope (in the collapsar scenario), the apparent structure depends on how relativistic beaming effects shape the emission from each part of the jet. From an observational point of view, it is the apparent structure that determines the isotropic equivalent energy, the observed luminosity function and the like; from a theoretical point of view, one would like to reconstruct the intrinsic structure to find out \eg the actual energy emitted by the jet and, through the efficiency, the total energy (kinetic plus internal and possibly magnetic) of the jet itself; thus, a clear distinction between the two, and some insight on their interdependence, are to be worked out.

The importance of such a distinction was partly pointed out in a remarkable work by \cite{2006AIPC..836..117G}, but their study assumed a constant bulk Lorentz factor profile $\Gamma(\theta)=\Gamma$. Few works to date (as far as we know) assign a variable Lorentz factor profile $\Gamma(\theta)$ to the jet (\eg \cite{kumar-sj-afterglow-2003,2013MNRAS.428.2430L}) and none reports predictions about the apparent structure of a GRB jet within such a model.

\vspace{10pt}
The structure of the article is the following:
\begin{enumerate}
 \item in \S\ref{sec:intrinsic} and \S\ref{subsec:apparent-examples} we give a rigorous definition of intrinsic and apparent structure, and we make some examples to clarify the concepts;
 \item in \S\ref{subsec:formulas} we introduce two formulas to compute the apparent structure and the spectrum of a SJ given its intrinsic structure; in the following subsection, we compare the predictions of our formulas with previous treatments and show that they are consistent;
 \item in what follows next, we analyze the particular case of a Gaussian intrinsic structure, showing that (under the assumption that also the Lorentz factor has a Gaussian profile) its apparent structure is not Gaussian, but rather it is quite well described by a power law; we then show that a Gaussian quasi-universal jet model, with very reasonable parameter values, is consistent with recent constraints from the observed luminosity function of GRBs;
 \item we show that the model is consistent with the Amati correlation being a viewing angle effect;
 \item in the Appendix we give detailed derivations of the formulas presented in \S \ref{subsec:formulas}.
\end{enumerate}

\section{Intrinsic Structure}
\label{sec:intrinsic}
Following \citet{Rossi-structured2002} and \citet{zhang-universalconfig2002} we define the \textit{intrinsic} structure of the jet as follows:
\begin{itemize}
 \item we set up a spherical coordinate system with the central engine at its origin and the jet directed along the z axis;
 \item we define the function $\epsilon(\theta)$ as the energy \textit{emitted} (during the prompt emission) by the portion of the jet comprised between $\theta$ and $\theta + d\theta$, divided by the corresponding solid angle, \ie $\epsilon(\theta)\equiv \eta\, dE(\theta)/2\pi \sin\theta\,d\theta$, where $dE$ here stands for the total energy (kinetic plus internal and possibly that of the magnetic field) of the jet portion, and $\eta$ is the prompt emission efficiency, which might as well depend on $\theta$;
 \item we assign a Lorentz factor $\Gamma(\theta)$ to the emitting material comprised between $\theta$ and $\theta + d\theta$ during the prompt emission.
\end{itemize}
The functions $\epsilon(\theta)$ and $\Gamma(\theta)$ then define what we call the \textit{intrinsic} structure of the jet.

\section{Apparent Structure}
\subsection{Definition}

\begin{figure}
 \begin{center}
 \includegraphics{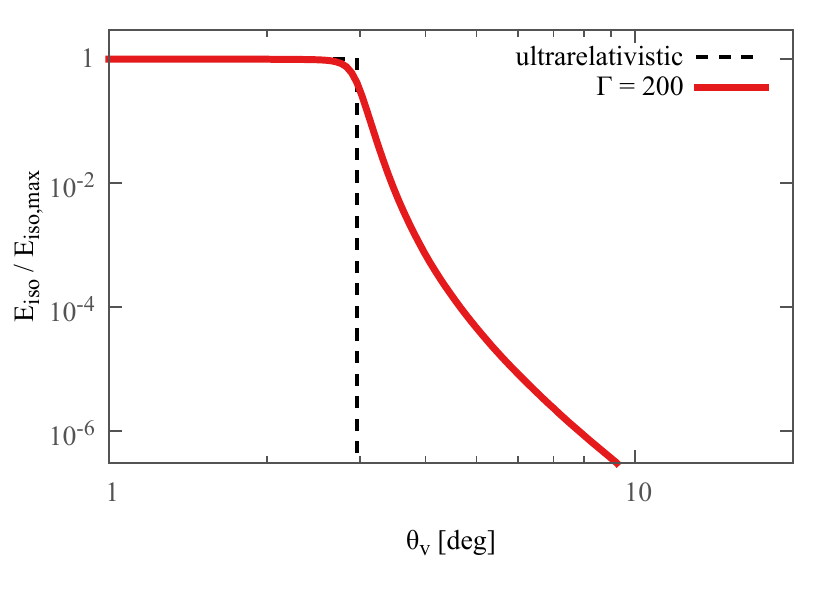}
 \caption{\label{fig:uniform-app-struct}Example apparent structure of a uniform jet in the ultrarelativistic limit (black dashed line) and for $\Gamma = 200$ (red solid line). The isotropic equivalent energy is normalized to its maximum value, corresponding to the jet observed on-axis. A jet semiaperture $\tj=3^\circ$ is assumed.}
 \end{center}
\end{figure}

\label{subsec:apparent-examples}
We introduce here our notion of \textit{apparent} structure. Let $\tv$ be the viewing angle of an observer looking at a GRB jet, \ie the angle between the jet axis and the line of sight. We call ``apparent structure'' the function $\Eiso(\tv)$, namely the isotropic equivalent energy inferred by the observer as a function of $\tv$. For the sake of clarity, let us apply this definition to some examples:
\begin{enumerate}
 \item an isotropic explosion, defined by $\epsilon(\theta)=\epsilon\;\;\forall \theta\in[0,\pi]$, would clearly have 
 \begin{equation}
 \Eiso = 4\pi\,\epsilon
 \end{equation}
 for all viewing angles;

 \item the ``classical'' uniform (``top-hat'') GRB jet has
\begin{equation}
\label{eq:epsilon-uniform}
  \epsilon(\theta) = \left\lbrace\begin{array}{lr}
   \epsilon & \theta<\tj \\
   0 & \theta \geq \tj \\
  \end{array}\right.\\
\end{equation}
 and
\begin{equation}
\label{eq:gamma-uniform}
  \Gamma(\theta) = \left\lbrace\begin{array}{lr}
   \Gamma & \theta<\tj \\
   1 & \theta \geq \tj \\
  \end{array}\right.\\
\end{equation}
 
 In the ultrarelativistic limit ($\Gamma \to \infty$) the uniform jet is indistinguishable from an isotropic explosion as long as $\tv < \tj$, because the relativistic beaming prevents \citep{rhoads-balloon97} the observer from ``seeing'' anything not expanding exactly along the line of sight\footnote{The implicit assumption is that the jet expansion is purely radial with respect to the central engine.}. For the same reason, the GRB is always undetected if $\tv > \tj$. In other words, the apparent structure (dashed black line in Fig.~\ref{fig:uniform-app-struct}) is
 \begin{equation}
  \Eiso(\tv) = \left\lbrace\begin{array}{lr}
   4\pi\,\epsilon & \tv<\tj \\
   0 & \tv \geq \tj \\
  \end{array}\right.\\
  \label{eq:app-struct-unif-ur}
 \end{equation}

 This ultrarelativistic, uniform jet picture is often used in theoretical works about GRBs;
 
 \item relaxing the ultrarelativistic assumption, one must in principle take into account the contribution to the observed flux coming from the whole emitting volume of the jet (the result of such calculation for the uniform jet is usually dubbed ``off-axis jet model'', e.g. \citet{2003ApJ...594L..79Y,2004ApJ...614L..13E,2006MNRAS.372.1699G,2006ApJ...645..436D}). For the uniform jet, the resulting apparent structure $\Eiso(\tv)$ has been computed numerically by many authors and it differs from Eq. \ref{eq:app-struct-unif-ur} in that the transition from the ``on-axis'' ($\tv < \tj$) to the ``off-axis'' ($\tv > \tj$) regime is obviously smoother, and a non-zero energy is received from the observer even at large viewing angles, since the radiation is not 100 per cent forward-beamed (red solid line in Fig.~\ref{fig:uniform-app-struct}).

\end{enumerate}

\subsection{A general formula for the apparent structure of a jet}
\label{subsec:formulas}

In the appendix we derive a formula to calculate the apparent structure $\Eiso(\tv)$ of a jet with a given (axisymmetric) intrinsic structure $\left\lbrace\epsilon(\theta),\Gamma(\theta)\right\rbrace$. It is valid under the assumptions that the emission comes from a geometrically and optically thin volume whose surface does not change significantly during the emission. According to our derivation, such apparent structure is given by

\begin{equation}
 \Eiso(\tv) = \int \dfrac{\delta^3(\theta,\phi,\tv)}{\Gamma(\theta)}\,\epsilon(\theta)\,d\Omega
 \label{eq:app-struct-complete}
\end{equation}
where $\tv$ is the angle between the line of sight and the jet axis, and $\delta$ is the relativistic Doppler factor. Here $\Eiso$ is understood as $4\pi\,d_L^2/(1+z)$ times the \textit{bolometric} fluence measured at the Earth ($d_L$ is the luminosity distance). A formula to calculate the observed time integrated spectrum under the same set of assumptions is also derived in the appendix (Eq. \ref{eq:spectrum-complete}). It reads
\begin{equation}
 \mathcal{F}(\nu,\tv) = \dfrac{1+z}{4\pi\,d_L^2\,}\int \dfrac{\delta^2(\theta,\phi,\tv)}{\Gamma(\theta)}\,\dfrac{f(x,\vec\alpha)}{\nu_0'f_{\vec\alpha}}\,\epsilon(\theta)\,d\Omega
 \label{eq:spectrum-for-2-branch-power-law-intensity}
\end{equation}
where we have set $x=(1+z)\nu/(\delta\nu_0')$ for neatness. Here  $f(x,\vec\alpha)$ is a dimensionless function which defines the comoving spectral shape, which can depend on an array $\vec\alpha$ of parameters (see the Appendix for more details on its definition), $\nu_0'$ is some typical frequency of the comoving spectrum, and
\begin{equation}
 f_{\vec\alpha} = \int_0^\infty f(x,\vec\alpha)\,dx
\end{equation}
Formula \ref{eq:spectrum-for-2-branch-power-law-intensity} can be used to compute the isotropic equivalent energy in a specific band, by using 
\begin{equation}
 E_{\mathrm{iso},[\nu_1,\nu_2]}(\tv) = \dfrac{4\pi\,d_L^2}{1+z}\int_{\nu_1/1+z}^{\nu_2/1+z} \mathcal{F}(\nu,\tv)d\nu
 \label{eq:eiso-in-band}
\end{equation}

\subsection{Comparison with previous studies}

\begin{figure}
 \includegraphics{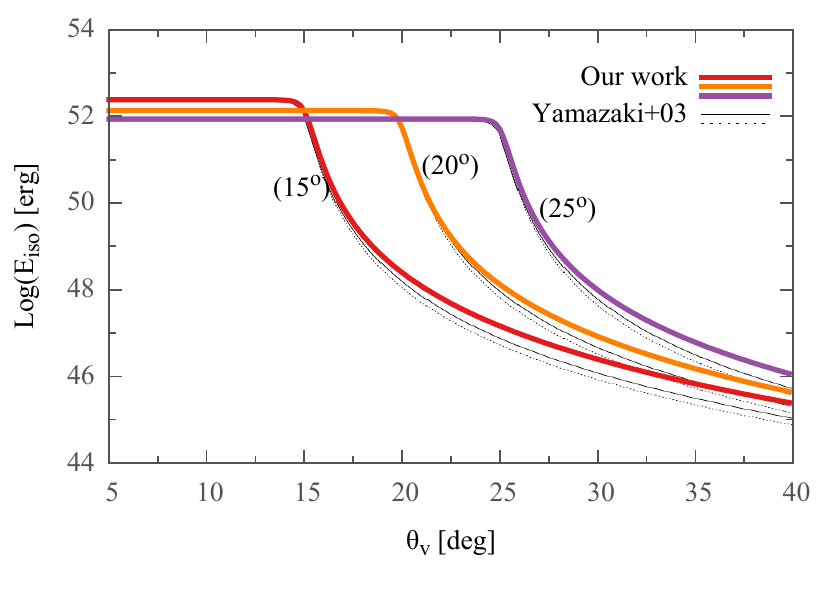}
 \caption{Apparent structures (Eq. \ref{eq:app-struct-complete}) of three uniform jets with $\Gamma=100$ and $\tj = 15^\circ$, $20^\circ$ and $30^\circ$ respectively (solid colored lines). The results of \citet{2003ApJ...594L..79Y} are shown for comparison (solid and dotted thin black lines). Our curves are normalized with the same prescriptions as in \citet{2003ApJ...594L..79Y}.}
 \label{fig:comparison-yamazaki}
\end{figure}

As a consistency check, we test our approach assuming a uniform jet structure (eqs. \ref{eq:epsilon-uniform} and \ref{eq:gamma-uniform}) and compare the results with the off-axis models of \citet[Y03 hereafter]{2003ApJ...594L..79Y}, \citet[E04 hereafter]{2004ApJ...614L..13E}  and \citet[G06 hereafter]{2006MNRAS.372.1699G}.
\begin{enumerate}
 \item the comparison with Y03 is obtained by using Eq. \ref{eq:eiso-in-band}, assuming the same redshift, comoving spectral shape, normalization, and Lorentz factor as in Y03. In Fig.~\ref{fig:comparison-yamazaki} we show our results together with those of Y03. Apparently, the model used in Y03 (thin black lines) slightly underestimates the off-axis isotropic equivalent energy with respect to ours (colored solid lines);
 \item the comparison with E04 is straightforward: the integrand of equation 3 of their work, which is used to calculate $\Eiso(\tv)$, is proportional to $\delta^3$, and the same holds for our Eq. \ref{eq:app-struct-complete} in the uniform jet case. The resulting apparent structures are thus the same up to a multiplicative constant;
 \item similarly, the integrand of equation 2 of G06, which is used to calculate the observed time-integrated spectrum, is proportional to $\delta^2$. Again, the same holds for our Eq. \ref{eq:spectrum-for-2-branch-power-law-intensity} in the uniform jet case. Integration of either equation over all frequencies gives an additional $\delta$ factor so that, as in the previous case, the resulting apparent structures are the same up to a multiplicative constant.
 \end{enumerate}
 We conclude that our model is reasonably consistent with previous studies on the uniform jet model, and it has the advantage that it can be applied to the SJ case using the definition of $\epsilon(\theta)$ commonly found in the literature.

\subsection{Application to power law and Gaussian jet models}

\begin{figure}
 \includegraphics{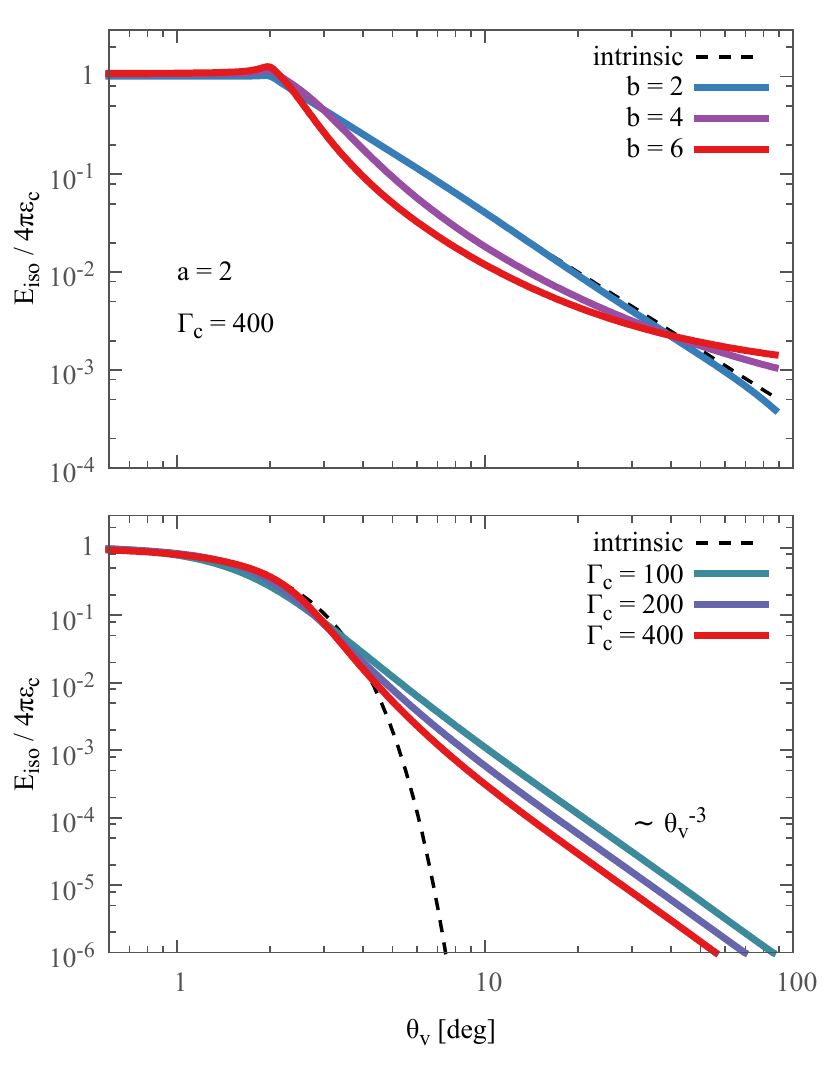}
 \caption{Apparent structures, according to Eq. \ref{eq:app-struct-complete}, of three power law (upper panel) and three Gaussian (lower panel) jet models. All power law models have $\tc = 2^o$, $\Gamma_c = 400$, $a=2$ and different values of the $b$ parameter, listed in the legend. The Gaussian models have $\tc = 2^o$ and different values of $\Gamma_c$, listed. The corresponding intrinsic energy structures (understood as $4\pi\epsilon(\tv)$) are shown (black dashed lines). The apparent structures in the Gaussian case decrease as $\Eiso\sim \tv^{-3}$ at large viewing angles, regardless of the core Lorentz factor $\Gamma_c$ and the jet typical angular size $\tc$.}
 \label{fig:app-struct-plaw-gaus}
\end{figure}

Here we want to show how the apparent structure of a SJ depends on the energy profile $\epsilon(\theta)$ and especially on the Lorentz factor profile $\Gamma(\theta)$: the latter has been assumed constant \citep{2006AIPC..836..117G} or its role has been deemed secondary \citep{Rossi-structured2002,zhang-universalconfig2002} in many preceding works. 

Fig.~\ref{fig:app-struct-plaw-gaus} shows the computed apparent structures of three power law jet models (upper panel) and three Gaussian jet models (lower panel). The intrinsic structures are defined following \citet{kumar-sj-afterglow-2003} as
\begin{equation}
\begin{array}{l}\epsilon(\theta)=\left\lbrace\begin{array}{lr}
					      \epsilon_c & \theta\leq\tc\\
					      \epsilon_c\left(\theta/\tc\right)^{-a} & \theta>\tc\\
					      \end{array}\right.\\
\rule{0pt}{20pt}                  
                  \Gamma(\theta)=\left\lbrace\begin{array}{lr}
                                              \Gamma_c & \theta\leq\tc\\
                                              1+(\Gamma_c-1)\left(\theta/\tc\right)^{-b} & \theta>\tc\\
                                             \end{array}\right.
\end{array}
\end{equation}
and
\begin{equation}
 \begin{array}{l}
  \epsilon(\theta) = \epsilon_c\;e^{-\left(\theta/\tc\right)^2}\\
  \Gamma(\theta) = 1 + (\Gamma_c-1)\;e^{-\left(\theta/\tc\right)^2}\\
 \end{array}
 \label{eq:intrinsic-structure-gaussian}
\end{equation}
for the power law and Gaussian jet model respectively. In both cases, the $\tc$ parameter represents the typical angular scale of the intrinsic structure, \ie the angle within which most of the jet energy is contained. Inspection of Fig.~\ref{fig:app-struct-plaw-gaus} shows that \textit{the more the Lorentz factor varies, the less the apparent structure mimics the underlying intrinsic structure}. The Gaussian jet model, in particular, displays an apparent structure which is quite well described by a power law with a slope around $-3$, plus a roughly constant core.

 \subsection{Reformulation of the Gaussian intrinsic structure}
 \label{sec:thetagamma}
 
 \begin{figure}
  \includegraphics{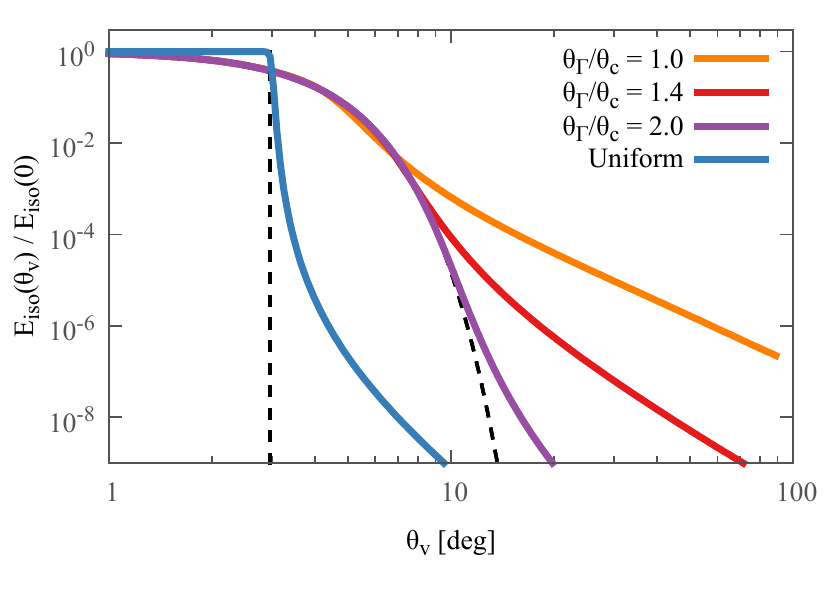}
  \caption{\label{fig:unif-gaus-comparison}Apparent structures of three Gaussian jets with different values of $\tgtc$ (reported near each line), together with the apparent structure of a uniform jet (blue solid line). The corresponding intrinsic structures are also shown (dashed lines). The Gaussian jets have $\tc = 3^\circ$ and $\Gamma_c = 400$; the uniform jet has $\tj=3^\circ$ and $\Gamma = 400$.}
 \end{figure}

 In the Gaussian case, the slope of the power law tail of the apparent structure at large viewing angles (see Fig.~\ref{fig:app-struct-plaw-gaus}) is almost unaffected by changes in the two parameters $\tc$ and $\Gamma_c$, and it remains between $-3$ and $-4$ for reasonable values of these parameters. A different slope can be achieved by modifying Eq. \ref{eq:intrinsic-structure-gaussian} as follows
\begin{equation}
 \begin{array}{l}
  \epsilon(\theta) = \epsilon_c\;e^{-\left(\theta/\tc\right)^2}\\
  \Gamma(\theta) = 1 + (\Gamma_c-1)\;e^{-\left(\theta/\theta_\Gamma\right)^2}\\
 \end{array}
 \label{eq:intrinsic-structure-gaussian-modified}
\end{equation}
where $\theta_\Gamma$ is a new parameter that allows for the Lorentz factor structure and intrinsic energy structure to fall off over different angular scales. In principle $\tc$ might differ from $\theta_\Gamma$ for the following reason:  $\epsilon(\theta)$ is related to the energy density $u = \rho c^2 + p + u_B$ (where $u_B$ is the magnetic energy density) of the jet according to
\begin{equation}
 \epsilon(\theta)\equiv\eta\dfrac{dE}{d\Omega}(\theta)=\eta\,\dfrac{4\pi R^2(\theta)\,\Delta(\theta)}{4\pi}\,u(\theta)
\end{equation} 
where $\Delta(\theta)$ is the width of the emitting volume and $R(\theta)$ defines its surface. The energy density $u$ is related to the comoving one by $u(\theta)=\Gamma^2(\theta)u'(\theta)$. Let us take the simplest picture as an example: (a) let the emitting volume be a portion of a spherical shell, with fixed width $\Delta$ and radius $R$; (b) let the efficiency $\eta$ be the same at all angles. One then gets $\epsilon(\theta)\propto u(\theta) = \Gamma^2(\theta)u'(\theta)$. If $u'$ is constant, this implies $\theta_\Gamma = \sqrt{2}\tc$. The efficiency, geometry and energy density all play a role in determining the ratio $\tgtc$. This ratio is the main parameter affecting the slope of the power law tail of the apparent structure\footnote{Let us remark that to get a Gaussian \textit{apparent} structure, as assumed \eg in \citet{2004ApJ...601L.119Z} and \citet{2005ApJ...621..875D}, one needs $\theta_\Gamma\gg\tc$, \ie the Lorentz factor should remain very high while the energy per unit solid angle falls off exponentially.}. Figure~\ref{fig:unif-gaus-comparison} shows the apparent structure of the Gaussian jet for three values of $\tgtc$, along with the uniform jet for comparison.

\section{The Gaussian SJ as a quasi-universal jet model}

\begin{table*}
\caption{\label{tab:quasi-univ-parameters}Set of parameters defining our quasi-universal Gaussian SJ.}
\begin{tabular}{llll}
\hline
Parameter & Value & Comment & Limits$^a$ \\
\hline
$  4\pi\epsilon_c $&$ 3\times 10^{53}$ erg& Needed to match the break in the observed LF. & $\sigma_{\log\epsilon_c}\lesssim 1\dex$ \\
$  t $&$ 22s$& Mode of the observed $T_{90}$ divided by average redshift $\left<1+z\right>$. & $0.35\dex\lesssim\sigma_{\log\Liso}\lesssim 0.78\dex$ \\
$  \Gamma_c $&$ 800$ & Highest $\Gamma$ inferred from the onset of an afterglow light curve. & $\Gamma_c\gtrsim 100$ \\
$  \tc $&$ 3^\circ$ & Gives a total energy of the same order of the break out energy. & $3^\circ \lesssim \tc \lesssim 5^\circ$ \\
$  \tgtc $&$ 1$ & Reasonable if the density of the jet core is less than that of the wings. & $0.5\lesssim\tgtc\lesssim \sqrt{2}$\\
\hline
\end{tabular}

\flushleft $^a$ all the limits are discussed within the text.
\end{table*}

\subsection{Luminosity function}
\label{sec:lf}

\begin{figure}
  \includegraphics{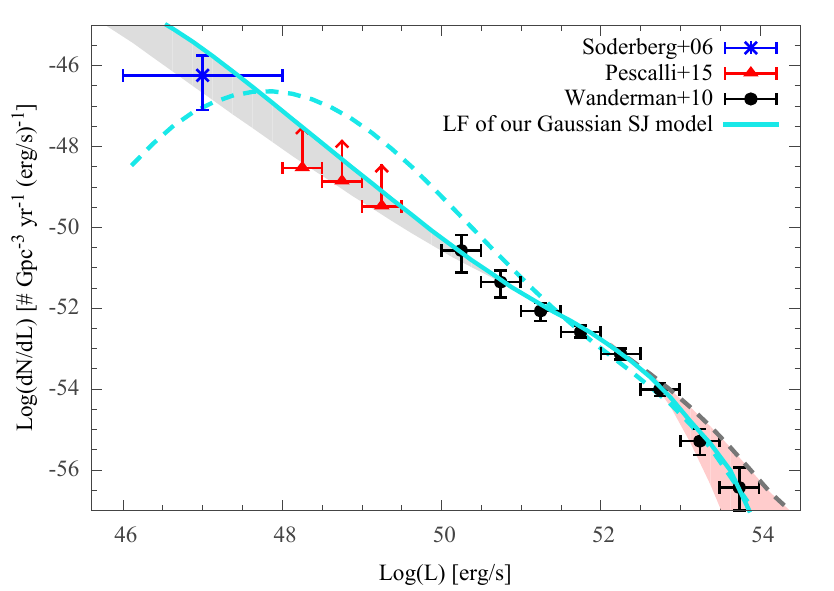}
  \caption{\label{fig:lf-gaus}The light blue line represents the luminosity function of our fiducial model (see Table~\ref{tab:quasi-univ-parameters}). There are two shaded areas obtained by varying one parameter and leaving the others fixed: the gray one refers to $1\leq\tgtc\leq\sqrt{2}$, while the pink one refers to $\sigma=\sqrt{\sigma_{\log t}^2+\sigma_{\log\epsilon_c}^2}$ between $0.35\dex$ and $0.78\dex$. The dashed gray line (visible on the bottom right corner) is the LF for $\tc=5^\circ$, while the dashed light blue line is the LF for $\tgtc = 0.5$, both with all other parameters fixed. The data points are the same as in \citet{2015MNRAS.447.1911P}, and are partly based on previous works by \citet{2006ApJ...638..930S} and \citet{2010MNRAS.406.1944W}. Red points are lower limits.}
 \end{figure}

The luminosity function of GRBs clearly depends upon their apparent structure rather than on the intrinsic one. We \citep{2015MNRAS.447.1911P} recently showed that a quasi-universal jet with a power law \textit{apparent} structure is consistent with the observed luminosity function (LF) of Long GRBs: according to our analysis above, the corresponding intrinsic structure might then be Gaussian. In order to test the possibility that a Gaussian quasi-universal SJ is compatible with the Long GRB LF, we first need a way to relate $\Eiso$ and the isotropic equivalent luminosity $\Liso$. The simplest approach is to define a rest frame duration of the burst $t$ and to assume a triangular shape for the prompt emission light curve. One then has 
\begin{equation}
 \Liso\sim 2\times \Eiso/t
 \label{eq:liso-from-eiso}
\end{equation}

To define our candidate quasi-universal Gaussian SJ we need a set of typical values of the model parameters. Here is an educated guess, based on heuristic arguments:
\begin{enumerate}
\item we define the typical rest frame duration $\left<t\right>$ as the mode of the observed prompt emission time distribution $\left<T_{90}\right>\approx 70 s$ \citep{2011ApJS..195....2S} divided by the average Long GRB redshift $\left<1+z\right>\approx 3.14$ \citep{2012ApJ...756..187H}, obtaining $\left<t\right>\approx 22 s$;
 \item the Long GRB luminosity function \citep{2010MNRAS.406.1944W} breaks around $L_{*} \approx 3\times 10^{52}$ erg/s: in the SJ picture, this luminosity corresponds to the typical GRB seen on-axis. By Eq. \ref{eq:liso-from-eiso} such GRB has an on-axis isotropic equivalent energy $\Eiso(\tv=0)\approx 3\times 10^{53}$ erg, or equivalently $\left<\epsilon_c\right> \approx 3\times 10^{53}\,\mathrm{erg}/4\pi \approx 2.4\times 10^{52}\, \mathrm{erg/sr}$. Since the highest measured $\Eiso$ so far is approximately $5\times10^{54}$ erg \citep[\eg GRB 080916C -][]{2010MNRAS.403..926G}, the $\epsilon_c$ parameter requires some dispersion to accommodate the observations;
 \item the total emitted energy\footnote{Some authors assume two equal, oppositely directed jets. To avoid confusion, we stress that we refer here to a single jet.} (during the prompt) is
 \begin{equation}
  E_{\gamma} = 2\pi\int_{0}^{\pi/2}\epsilon(\theta)\sin\theta\,d\theta \approx \pi\epsilon_c\tc^2
 \end{equation}
 According to \citet{2014MNRAS.445..528K}, a typical jet employs around $10^{51}$ erg to break out the envelope of the star in the collapsar scenario: requiring the remaining energy $E_{\mathrm{jet}}=E_{\gamma}/\eta$ to be at least of the same order, assuming an efficiency $\eta = 0.2$ we obtain a lower limit on the jet angular scale
 \begin{equation}
\tc\gtrsim 3^\circ
 \end{equation}
A jet with an aperture smaller than this must have lost more than half of its initial energy in the excavation of its channel through the star envelope;
\item some mixing is likely to occur between the jet borders and the stellar envelope \citep{Rossi-structured2002}, and indeed simulations indicate (e.g. \citet{morsony-grb-simula-2007}) that the jet plasma density increases with the distance from the axis. In the simplest case outlined in \S\ref{sec:thetagamma}, this suggests a ratio $\tgtc \lesssim \sqrt{2}$. We thus take $\tgtc = 1$ for simplicity;
\item the exact value of $\Gamma_c$ has little effect on the apparent structure of the Gaussian jet (as long as it is $\gtrsim 100$), so it is a secondary parameter for what concerns the LF. By the way, let us remark that hydrodynamic simulations by \citet{kumar-sj-afterglow-2003} suggest that for a Gaussian SJ the afterglow onset time is soonest for $\tv=0$, in which case it is the same as that of a uniform jet with $\Gamma=\Gamma_c$. Within the SJ picture, this indicates that the highest Lorentz factors inferred so far (those of GRBs detected in the GeV energy range by Fermi-LAT, see \citet{ghirlanda-comoving-2012}) are a measure of the core Lorentz factor $\Gamma_c$ of the underlying jets. We therefore give this parameter the rather high value $\Gamma_c=800$.
\end{enumerate}
The set of typical parameter values that we induced from these arguments is given in Table~\ref{tab:quasi-univ-parameters}. As stated in point (ii) above, some dispersion in $\epsilon_c$ around its typical value is necessary to match the observations. Similarly, observations show that the rest frame emission time $t\sim T_{90}/1+z$ is certainly not universal \citep[e.g.][]{2011ApJS..195....2S}: in the next section we will discuss how to handle these two parameters and their dispersion.

Before moving on, let us note that we are assuming no evolution of the typical values with redshift: this might well be a rough approximation, since the overall progenitor properties may vary with redshift. Nevertheless, to keep the discussion as simple as possible, we neglect this aspect and assume that the quasi-universal jet is the same at all times in the past.

\subsubsection{Dispersion on emission time and jet total energy}
\label{sec:dispersions}

As a starting point, we assume a lognormal distribution for both the intrinsic duration $t$ and the core energy parameter $\epsilon_c$ with a total dispersion parameter $\sigma_{\log\Liso} = \sqrt{\sigma_{\log t}^2+\sigma_{\log\epsilon_c}^2}$ (the dispersion on the luminosity is affected the same way by the dispersion on $t$ and on $\epsilon_c$ because of Eq. \ref{eq:liso-from-eiso}). Figure~\ref{fig:lf-gaus} shows the LF of the model assuming $\sigma_{\log\Liso} = 0.56\dex$ (light blue solid line), together with the data used in \citet{2015MNRAS.447.1911P}. This is the value of $\sigma_{\log\Liso}$ for which we find the best agreement between the theoretical LF and the observed one. Varying $\sigma_{\log\Liso}$ between $0.35\dex$ and $0.78\dex$ one still obtains LFs (pink shaded area in Fig.~\ref{fig:lf-gaus}) that lie within the error bars of the data points, so we take these values as estimates of the limits of this parameter\footnote{Let us remark that the large uncertainties in the observed LF make it not very constraining: as a matter of fact, the reduced Chi-squared of the model in Fig.~\ref{fig:lf-gaus} is formally $\tilde{\chi}^2\sim 10^{-1}$. Indeed, we have shown in \citet{2015MNRAS.447.1911P} that several jet models can reproduce the observed LF to date.}. 

In a realistic model, one might expect the core energy per unit solid angle $\epsilon_c$ to correlate with the emission time $t$, so a more rigorous approach is to interpret the parameter $\sigma_{\log\Liso}^2$ as the \textit{residual variance} of $\log \epsilon_c$ with respect to its (possible) correlation with $\log t$.
We have two limiting cases:

\begin{enumerate}
 \item in the ``worst'' case, a linear relation holds between $\epsilon_c$ and $t$, so that the dispersion on $t$ gives no contribution to the dispersion on $\Liso$. Let us take the standard deviation of $\log T_{90}$ (for Long GRBs observed by \textit{Swift}), which is around $0.57\dex$ \citep{2011ApJS..195....2S}, as an estimate of $\sigma_{\log t}$. The dispersion parameter $\sigma_{\log\epsilon_c}$ (and consequently the dispersion on the logarithm of the jet total energy) required in this case to reproduce the LF of Fig.~\ref{fig:lf-gaus} is then $\sim 1.13\dex$;
 \item at the other end, if $t$ and $\epsilon_c$ were independent, a $0.56\dex$ dispersion on $\log t$ only would be sufficient to reproduce the LF: in other words, a single universal value of the jet energy would be consistent with the LF (but not with the observed $\Eiso$ distribution, as noted in the preceding section).
\end{enumerate}
   As a result, Fig.~\ref{fig:lf-gaus} suggests that a quasi-universal value $\left<\epsilon_c\right>\approx 2.4\times 10^{52}\, \mathrm{erg/sr}$ (which corresponds to $E_{\mathrm{jet}} = 10^{51}\erg$ if $\tc = 3^\circ$ and $\eta=0.2$) with a dispersion of less than $1\dex$ (but not much less) is compatible with the considered observational constraints. This goes well along with the fact that the progenitor (a former Wolf-Rayet star) is expected to possess rather standard properties.
   
   At this point, a remark is necessary: we assumed that the distribution of $t$ is independent from the luminosity, which might be questionable. Indeed, some authors argue \citep{2007A&A...465....1D,2009MNRAS.392...91V,2011ApJ...739L..55B,arxiv-1503.00441,2015MNRAS.448..417B} that low luminosity, very long GRBs might represent a distinct population, possibly originating from a different progenitor. One of such GRBs is included in the low luminosity bin of Fig.~\ref{fig:lf-gaus}, namely GRB060218 (the only other burst in the bin is GRB980425, which lasted $\sim 40 s$). Based on such a distinction, one may suppose that there is an anti-correlation between luminosity and duration in the overall population. On the other hand, though, there is another subclass of GRBs \cite[the so called ``ultra-long'' GRBs, see][]{2014ApJ...790L..15P,2014ApJ...781...13L} which have durations around $10^4$ s, but are not underluminous: indeed, no clear correlation exists between luminosity and duration within today's samples. We thus conclude that, for our simple model, the assumption of a distribution of emission times which does not depend on the burst luminosity is acceptable.
   
\subsubsection{Limits on the ratio $\tgtc$}
While the dispersion $\sigma_{\log \Liso}$ affects the high luminosity end of the LF in Fig.~\ref{fig:lf-gaus}, the value of the ratio $\tgtc$ has its influence on the low luminosity end. As explained in \S\ref{sec:thetagamma}, different values of the ratio yield different slopes of the apparent structure. The steeper the slope (the higher $\tgtc$), the fainter the jet when it is seen under a 90 degrees viewing angle: this implies a lower limit on $\tgtc$ if we require the predicted LF to extend down to the lowest observed luminosities. For our model, this lower limit is $\tgtc \sim 0.5$ (light blue dashed line in Fig.~\ref{fig:lf-gaus}); increasing $\tgtc$ from $1$ to $\sqrt{2}$ one obtains the family of LFs spanning the gray shaded area in Fig.~\ref{fig:lf-gaus}. The highest value of $\tgtc$ consistent with the lower limits on the rate of intermediate luminosity GRBs (red points in Fig.~\ref{fig:lf-gaus}) is $\tgtc \sim \sqrt{2}$.

\subsubsection{Limits on the angular scale $\tc$}
The width of the angular scale $\tc$ impacts mainly on the high luminosity end of the LF. The wider $\tc$, the higher the probability of observing the jet within the core (\ie $\tv\leq\tc$), implying a higher rate of bursts with high observed luminosity. The dark grey dashed line in Fig.~\ref{fig:lf-gaus} shows how the LF would change if $\tc=5^\circ$ was assumed. A wider $\tc$ would lead to higher rates than those observed at high luminosities.

As explained in \S\ref{sec:lf}, the lower limit $\tc\sim 3^\circ$ is given by the requirement that the jet total energy $E_{\mathrm{jet}}=E_{\gamma}/\eta$ of the quasi-universal jet is at least $10^{51}$ erg, \ie of the same order of the energy previously spent by the jet to excavate its way through the stellar envelope \citep{2014MNRAS.445..528K}. This is not a strict requirement, since one can also have that most of the jet energy is spent prior to the break out (we stress that $E_{\mathrm{jet}}$ is the total jet energy \textit{after} the break out).

\subsection{Consistency with the Amati correlation}
\label{sec:amati}

Gamma Ray Bursts show a strong correlation \citep{Amati2002,0004-637X-620-1-355,2000ApJ...534..227L} between the peak of their $\nu F_{\nu}$ spectrum ($\Ep$) and the isotropic equivalent energy $E_{\rm iso}$, roughly
\begin{equation}
 E_{\rm peak}\propto E_{\rm iso}^{1/2}
\end{equation}
A similar correlation involves the isotropic equivalent luminosity $L_{\rm iso}$ \citep{2004ApJ...609..935Y}. 
These correlations have been extensively studied in the past years for (a) their possible physical implications on the GRB emission process (e.g. \citet{2005ApJ...628..847R,2005ApJ...635..481T,2006AIPC..836...91B,2006ApJ...652.1400R,2006ApJ...651..333T,2007A&A...469....1G,2007ApJ...666.1012T}) and on the GRB jet structure \citep{2004ApJ...606L..33Y,2004ApJ...614L..13E,0004-637X-620-1-355,2005ApJ...629L..13L}, (b) the possibility to use them to standardize GRB energetics and constrain the cosmological parameters \citep{2004ApJ...613L..13G,amati-cosmological-parameters-2008}. Besides, these correlations stimulated an intense debate on the possible impact of selection effects \citep{2005MNRAS.360L..73N,2005ApJ...627..319B,2007ApJ...671..656B,2009ApJ...694...76B,2009AIPC.1133..425S,2012ApJ...747..146K,2005MNRAS.361L..10G,2008MNRAS.384..599B,2008MNRAS.387..319G,2008MNRAS.391..639N,2009A&A...508..173A,2009ApJ...704.1405K,2012MNRAS.422.2553G,2013A&A...557A.100H}. Despite the wealth of papers, the spectral--energy correlations are still a hot subject in the field and no consensus on their physical interpretation has been reached yet. 

It has been proposed recently that the Amati correlation might be due to a sequence of bulk Lorentz factors, with more luminous GRBs having larger $\Gamma$ values, as suggested by the possible clustering of the GRB energetics when transformed in the comoving frame \citep{ghirlanda-faster-narrower-2013}.

\begin{figure}
  \includegraphics{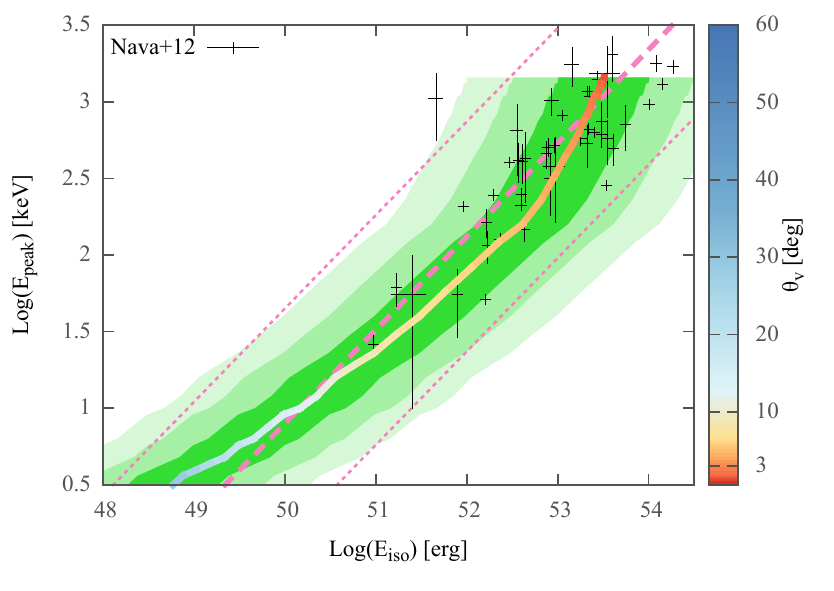}
  \caption{\label{fig:amati}The thick solid colored line represents a sequence of $\Eiso(\tv)$ and $\Ep(\tv)$ for a Gaussian SJ with the parameters given in Tab.\ref{tab:quasi-univ-parameters}. The color coding of the curve accounts for the viewing angles according to the scale defined on the right. Data points from the Swift BAT complete GRB sample \citep{2012MNRAS.421.1256N} are shown for comparison. The pink dashed lines represent the Amati correlation and the $3\sigma$ dispersion of the data points as computed by \citet{2012MNRAS.421.1256N}. The shaded area represents the portion of the plane spanned by the curves $\left\lbrace\Eiso(\tv),\Ep(\tv)\right\rbrace$ with $\log\epsilon_c$ varying within $\pm\sigma$ (darkest shade), $\pm2\sigma$ and $\pm3\sigma$ (lightest shade) respectively (we assumed $\sigma = 0.5\dex$).}
 \end{figure}

Here we wish to show that, within the quasi-universal Gaussian SJ defined in the preceding section, the Amati correlation can be explained as a viewing angle effect. First, we need to make some assumptions on the comoving spectrum emitted by the jet:

\begin{enumerate} 
 \item following \eg \citet{2003ApJ...594L..79Y,2006MNRAS.372.1699G} we assume a smoothly broken power law shape for the comoving spectrum. In the notation of Eq. \ref{eq:spectrum-for-2-branch-power-law-intensity}, this amounts to write
\begin{equation}
 f(x,a,b) = \dfrac{1}{x^a + x^b}
 \end{equation}
 We choose $a=-0.1$ and $b=1.3$, which are the typical low and high energy slopes of observed GRB energy spectra (\eg \citet{2011A&A...530A..21N}) (the results of this work do not change if we take different values of $a$ and $b$ within their typical dispersion);
\item we choose the peak frequency of the comoving spectrum to be $\nu_0'=1\keV$ at all points in the jet. This is motivated by the finding \citep{ghirlanda-comoving-2012} that the peak of GRB spectra seems to cluster in the comoving frame\footnote{One could also take $\nu_0'=\nu_0'(\theta)$, \ie give a structure to the peak of the comoving spectrum, and this would certainly lead to different results. Since we have no argument to prefer such a choice, we limit ourselves to the simpler case discussed in the text.}. To choose a different value for $\nu_0'$ amounts to move vertically the colored curve in Fig.~\ref{fig:amati}.

\end{enumerate}
 With the above assumptions, we used Eq. \ref{eq:spectrum-for-2-branch-power-law-intensity} to calculate the observed spectrum and Eq. \ref{eq:eiso-in-band} to compute the corresponding $\Eiso$ in the $[1\keV-10^4\keV]$ band. The solid colored line in Fig.~\ref{fig:amati} represents the resulting sequence of $\Eiso(\tv)$ and $\Ep(\tv)$ for our Gaussian SJ with the parameters given in Tab. \ref{tab:quasi-univ-parameters}. Data points (for comparison) are bursts from the Swift BAT complete sample \citep{2012ApJ...749...68S} which is flux limited and 97 per cent complete in redshift; the pink dashed lines represent the Amati correlation and the $3\sigma$ dispersion of the data points as computed by \citet{2012MNRAS.421.1256N} using the same sample. Changing the value of $\epsilon_c$ (a dispersion of $\epsilon_c$ is necessary to reproduce the LF, as discussed in the preceding section) amounts here to moving horizontally the colored curve: assuming $\sigma_{\log\epsilon_c}=0.5\dex$, we colored the portions of the plane spanned by the curve when $\log\epsilon_c$ varies within $\pm\sigma$ (darkest shade), $\pm2\sigma$ and $\pm3\sigma$ (lightest shade) respectively.
 
 We conclude that this model is compatible with the interpretation of the Amati correlation as a sequence of viewing angles, with the dispersion of the observed points being due to the intrinsic dispersion in the total jet energy and possibly in the peak of the comoving spectrum.

\section{Discussion and Conclusions}

To some extent, structure is an inevitable feature of GRB jets, since a uniform jet with discontinuous edges is clearly unphysical: indeed, a Gaussian jet structure has been often described as a more realistic version of the uniform jet structure (\eg \citet{zhang-universalconfig2002}) and it is expected to reproduce most of the features of the latter when the viewing angle is small enough \citep{kumar-sj-afterglow-2003}. 

In this view, our Gaussian SJ model is to be understood as a proxy for a jet in which \textit{both} the emissivity \textit{and the Lorentz factor} decrease rapidly away from the axis. Our results then indicate that the simplified ``ultrarelativistic uniform (top-hat)'' jet model, while very useful, likely predicts too little off-axis energy emission with respect to any more realistic counterpart. This might be a clue to the still missing observation of ``orphan afterglows'': the jet can be narrow (which is a necessary condition, in the collapsar scenario, for the jet to break out the progenitor star envelope without wasting too much energy), and still be visible at viewing angles larger than its typical angular dimension\footnote{The reduction of the expected number of orphan afterglows is a common feature of SJ models, see \citet{2008MNRAS.390..675R}.}, contrary to the uniform jet model \citep{1998ApJ...509L..85P}. This affects also the interpretation of the luminosity function of GRBs: the high rate of underluminous events and the wide range of observed luminosities (from $10^{47}$ erg/s up to $10^{54}$ erg/s) are readily explained without invoking the existence of different burst populations \citep{2015MNRAS.447.1911P}.

How standard can be the jet of GRBs in this picture?
The observed break in the luminosity function of Long GRBs at $L_{*}\sim 3\times 10^{52}$ erg/s \citep{2010MNRAS.406.1944W} sets a natural scale for the luminosity. Within the quasi-universal structured jet hypothesis, this luminosity corresponds to the typical jet seen on-axis. 
The on-axis $E_{iso}$ of the typical jet is thus $E_{*}\sim 3\times 10^{53}$ erg (the jet total energy $E_{\mathrm{jet}} = E_{\gamma}/\eta \approx 2\pi\tc^2\epsilon_c$ depends on the angular dimension $\tc$ of the jet, and it is $E_{\mathrm{jet}}\approx 10^{51}$ erg for $\tc = 3^\circ$). The fact that we do observe more energetic GRBs (up to $E_{iso}\sim 5\times 10^{54}$ erg) means that some dispersion in the jet energy is necessary to account for it. On the other hand, a dispersion of around $0.6 \dex$ in the maximum luminosity $L_{*}$ is needed to reproduce the luminosity function. As discussed in \S\ref{sec:dispersions}, this limits the dispersion on the jet total energy below $1\dex$. Figure~\ref{fig:amati} shows that $0.5\dex$ is enough to account for the dispersion in the Amati correlation. We can conclude that the jet of GRBs in this picture can be rather standard, with a total energy that lies within a factor of 10 from the typical $E_{\mathrm{jet}}\approx 10^{51}$~erg in most cases. This is exactly what one would expect by the association of GRBs with supernovae Ib/c, since the latter should have rather standard progenitors.

The explanation of the Amati correlation as a viewing angle effect has been proposed several times in the past, within a variety of jet models (\eg \citet{2004ApJ...606L..33Y,2004ApJ...614L..13E,0004-637X-620-1-355,2005ApJ...629L..13L,2006AIPC..836..117G}; recently also in a photospheric emission model by \citet{2013ApJ...765..103L}, where the jet structure is computed through relativistic hydrodynamical simulations of the jet break out). The result presented in \S\ref{sec:amati} shows that this interpretation is possible also in our simple model, at least within the discussed assumptions. In this view, the dispersion of the observed correlation is due to the intrinsic dispersion on the jet total energy (and possibly on the comoving peak energy).

We conclude by stressing (as noted in \S\ref{sec:lf}) that in this simplified model we neglected the possible evolution of the universal jet parameters with redshift, which might play an important role in determining the luminosity function and other observational features of the GRB population. We also assumed no correlation between emission time and luminosity, for the reasons explained in the last paragraph of \S\ref{sec:dispersions}.

\section*{Acknowledgements}
We thank the anonymous referee for pointing out some inaccuracies in the manuscript draft. We thank Davide Lazzati for patiently sharing his knowledge and understanding about jet break out simulations. We also thank Giacomo Bonnoli for his contribution in a discussion about statistics. G. Ghirlanda acknowledges PRIN - INAF 2011 for financial support.

\section*{References}

\footnotesize{
\bibliographystyle{mn2e}
\bibliography{journals,grb}}

\section{Appendix}

\subsection{Derivation of the formula for $\Eiso(\tv)$}

\begin{figure}
 \begin{center}
 \includegraphics{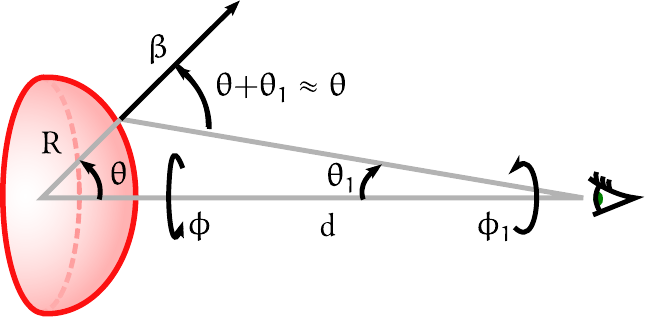}
 \caption{\label{fig:geom1}An uniformly expanding hemispherical shell emits radiation for a short time interval, during which its radius is nearly constant and equal to $R$. An observer is at a distance $d\gg R$. Each point of the hemisphere is moving in a different direction, thus contributing differently to the flux due to relativistic beaming.}
 \end{center}
\end{figure}

As a first step, consider an uniformly expanding hemispherical shell which emits radiation for a small time interval $\Delta t$, during which the radius does not vary appreciably (see Fig.~\ref{fig:geom1} for a sketch of the geometry). For definiteness, we set up two coordinate systems: the origin of the first (call it $K$) is the center of the hemisphere, and its $z$ axis lies on the line connecting this point to the observer. The spherical coordinates of this system will be referred to as $(r,\theta,\phi)$. The second system ($K_1$) is centered on the observer, and its $z_1$ axis coincides with $z$, but it is oppositely oriented. The corresponding spherical coordinates will be called $(r_1,\theta_1,\phi_1)$. If there is no significant absorption, the flux received by an observer at a distance $d$ is (neglecting cosmological corrections)

\begin{equation}
 F(\nu,t) = \dfrac{1}{d^2} \int_{\mathcal{S}(t)} d\phi_1 \, \sin\theta_1 d\theta_1 \, r_1^2 dr_1 \,\, j_\nu(\vec{r}_1, t-r_1/c)
\end{equation}
where $\mathcal{S}(t)$ is the equal arrival time surface for photons received at time $t$, and $j_\nu$ is the specific emissivity.
To obtain the corresponding fluence $\mathcal{F}(\nu)$, we must integrate over time. This allows us to ``rearrange'' the emission times so that we do not need to bother about the equal arrival time surfaces, \ie

\begin{equation}
\mathcal{F}(\nu) = \dfrac{1}{d^2} \int_{\Delta t} dt \int_{V(t)} d\phi_1 \, \sin\theta_1 d\theta_1 \, r_1^2 dr_1 \; j_\nu(\vec{r}_1,t) \\
\end{equation}
where $V(t)$ is the emitting volume at time $t$. 
Now we introduce some simplifications:
\begin{enumerate}
 \item the hemisphere is far away from the observer ($d \gg R$), so that we can safely set $r_1 \approx d$, $\sin\theta_1\approx \theta_1$ and $\theta_1 \approx \sin\theta\, R/d$;
\item since we integrate $\phi_1$ from $0$ to $2\pi$, we can equivalently integrate over $\phi$;
\item we assume that the emitting volume is geometrically thin, \ie it is a hemispherical shell of width $\Delta r$.
\end{enumerate}

Since $dr_1\cos\theta = dr$, we have
\begin{equation}
 \mathcal{F}(\nu) = \dfrac{R^2}{d^2} \int_{\Delta t} dt  \int_{0}^{\pi/2} \sin\theta\,d\theta\int_0^{2\pi}\,d\phi\int_{\Delta r} \;j_\nu(\vec{r},t)dr
\end{equation}
Now we use the relations $j_\nu = \delta^2\,j_{\nu'}'$ \citep{rybickilightman} where $\delta$ is the relativistic Doppler factor\footnote{We are implicitly assuming that the emissivity is isotropic in the comoving frame.} and primed quantities refer to the comoving frame, and $dr = \delta dr'$ to integrate over $\Delta r'$ and obtain

\begin{equation}
\mathcal{F}(\nu) = \dfrac{R^2}{d^2}\int_{\Delta t} dt  \int_{0}^{\pi/2} \sin\theta\,d\theta\int_0^{2\pi}d\phi\;\delta^3 I'_{\nu'}(\theta,\phi,t)
\label{eq:specific-fluence-given-intensity}
\end{equation}
where $I'_{\nu'}$ is the comoving specific intensity. Integrating over $dt=dt'/\delta$ and $d\nu=\delta d\nu'$ to get the bolometric fluence, we have
 \begin{equation}
 \mathcal{F} = \dfrac{R^2}{d^2} \int_{0}^{2\pi}\int_{0}^{\pi/2} \delta^3\, <I'>(\theta,\phi)\,\Delta t'\sin\theta d\theta\,d\phi
\label{eq:fluence-complete}
 \end{equation}
where  $<I'>$ is the average comoving intensity during the emission time $\Delta t'$. If the sphere emits uniformly, $<I'>$ does not depend on $\theta$ and $\phi$: in this case the integral is analytic, yielding
\begin{equation}
 \mathcal{F} = \dfrac{\pi\,R^2\Gamma\left(1+\beta\right)^2(2-\beta)}{d^2}\,<I'>\Delta t'
\end{equation}
By definition we have $\Eiso = 4\pi\,d^2 \mathcal{F}$, so we finally get
\begin{equation}
 \Eiso = 4\pi^2(1+\beta)^2(2-\beta) R^2\,\Gamma<I'>\Delta t'
\end{equation}
As stated in section \ref{subsec:apparent-examples}, in the ultrarelativistic limit the isotropic equivalent energy is indistinguishable from that of a spherical explosion, which yields $\Eiso=4\pi\epsilon$ by definition, thus in this case we have
\begin{equation}
 16\pi^2 R^2\,\Gamma<I'>\Delta t' = 4\pi\epsilon
\end{equation}
so that we can make the identification
\begin{equation}
 <I'>\Delta t' = \dfrac{\epsilon}{4\pi\,R^2\,\Gamma}
 \label{eq:correspondence-I-epsilon}
\end{equation}
Since the shell is geometrically thin, the intensity coming from a point $(\theta,\phi)$ is due only to the local emitting volume, so in this approximation we can think of Eq. \ref{eq:correspondence-I-epsilon} as a relation between local quantities, namely
\begin{equation}
 <I'>(\theta,\phi)\Delta t'(\theta,\phi) = \dfrac{\epsilon(\theta,\phi)}{4\pi\,R^2\,\Gamma(\theta,\phi)}
\end{equation}
We thus substitute this equivalence back into Eq. \ref{eq:fluence-complete} and multiply it by $4\pi\,d^2$, to get
\begin{equation}
 \Eiso = \int_{0}^{2\pi}\int_{0}^{\pi/2} \dfrac{\delta^3(\theta,\phi)}{\Gamma(\theta,\phi)}\,\epsilon(\theta,\phi) \,\sin\theta\,d\theta\,d\phi
 \label{eq:app-struct-complete-costheta}
\end{equation}
In words, this equation tells us how to weigh the contribution from each element of the hemisphere in order to take into account the relativistic effects and the local energy density $\epsilon(\theta,\phi)$.

As long as the expansion is purely radial, this equation holds for elements of any surface - in other words, setting $R=R(\theta,\phi)$ does not affect the validity of the derivation. In the case of a structured jet we have $\epsilon=\epsilon(\theta)$, $\Gamma=\Gamma(\theta)$ and, if the observer is off-axis, the Doppler factor $\delta$ must take into account the angle between the line of sight and the velocity of each point of the surface. A little geometry allows one to write
\begin{equation}
 \delta(\theta,\phi,\tv) = \dfrac{1}{\Gamma(\theta)\left[1-\beta(\theta)\cos\alpha(\theta,\phi,\tv)\right]}
\end{equation}
with
\begin{equation}
\cos\alpha(\theta,\phi,\tv) = \cos\theta\cos\tv + \sin\theta\sin\phi\sin\tv
\label{eq:costhetaprime}
\end{equation}
where we assumed (without loss of generality) that the line of sight lies on the $z-x$ plane\footnote{\label{footnote:costheta}Some authors prefer to set the coordinates so that the line of sight lies on the $y-z$ plane: in this case, one would have $$\cos\alpha(\theta,\phi,\tv) = \cos\theta\cos\tv + \sin\theta\cos\phi\sin\tv$$}.

Finally
\begin{equation}
 \Eiso(\tv) = \int \dfrac{\delta^3(\theta,\phi,\tv)}{\Gamma(\theta)}\,\epsilon(\theta) \,d\Omega
 \label{eq:app-struct-complete-appendix}
\end{equation}

Summarizing, the formula above gives the apparent structure of a jet, given its intrinsic structure (\ie $\epsilon(\theta)$ and $\Gamma(\theta)$), seen under a given viewing angle $\tv$, under the assumption that the emission comes from a thin, transparent volume, whose surface $R(\theta,\phi)$ does not change significantly during the emission.

\subsection{Derivation of the formula for $\mathcal{F}(\nu,\tv)$}
\label{sec:spectrum}
We can also derive the corresponding formula for the time integrated spectrum $\mathcal{F}(\nu)$ as a function of the viewing angle $\tv$.
First, let us write the comoving specific intensity $I'_{\nu'}$ as $(I_0'/\nu_0')\, f(\nu'/\nu_0',\vec{\alpha})$, where $I_0'$ is a normalization constant, $\nu_0'$ is some preferred frequency and $f(\nu'/\nu_0',\vec{\alpha})$ is a dimensionless function that defines the shape of the comoving spectrum, which depends on an array $\vec{\alpha}$ of parameters. Let us also call $f_{\vec{\alpha}}$ the integral of $f(x,\vec{\alpha})$ over all positive $x$'s. Then we rewrite Eq. \ref{eq:correspondence-I-epsilon} as
\begin{equation}
 I_0' =  \dfrac{\epsilon}{\Delta t' f_{\vec{\alpha}}\,4\pi\,R^2\,\Gamma}
 \label{eq:I0prime}
\end{equation}
Starting again from Eq. \ref{eq:specific-fluence-given-intensity}, taking into account the above definitions we end up with

\begin{equation}
 \mathcal{F}(\nu,\tv) = \dfrac{1+z}{4\pi\,d_L^2\,}\int \dfrac{\delta^2(\theta,\phi,\tv)}{\Gamma(\theta)}\,\dfrac{f(x,\vec{\alpha})}{\nu_0'f_{\vec{\alpha}}}\epsilon(\theta)\,d\Omega
\label{eq:spectrum-complete}
 \end{equation}
where the comoving spectral shape $f(x,\vec{\alpha})$ is to be evaluated at $x=(1+z)\nu/(\delta\nu_0')$.

\label{lastpage}
\end{document}